\documentclass[conference]{IEEEtran}
\IEEEoverridecommandlockouts

\usepackage{cite}
\usepackage{amsmath,amssymb,amsfonts}
\usepackage{algorithmic}
\usepackage{graphicx}
\usepackage{textcomp}
\usepackage{xcolor}

\begin{document}

\title{How long can you sleep? Idle Time System Inefficiencies and Opportunities}
\author{
\IEEEauthorblockN{Georgia Antoniou}
\IEEEauthorblockA{
\textit{University of Cyprus}\\
}
\and
\IEEEauthorblockN{Haris Volos}
\IEEEauthorblockA{
\textit{University of Cyprus}\\
}
\and
\IEEEauthorblockN{Jawad Haj Yahya}
\IEEEauthorblockA{
\textit{Rivos Inc.}\\
}
\and
\IEEEauthorblockN{Yiannakis Sazeides}
\IEEEauthorblockA{
\textit{University of Cyprus}\\
}
}
\maketitle

\begin{abstract}


This work introduces a model-based framework that reveals the idle opportunity of modern servers running latency-critical applications. Specifically, three queuing models, M/M/1, c$\times$M/M/1, and M/M/c, are used to estimate the theoretical idle time distribution at the CPU core and system (package) level. A comparison of the actual idleness of a real server and that from the theoretical models reveals significant missed opportunities to enter deep idle states. This inefficiency is attributed to the idle-governor inaccuracy and the high latency to transition to/from legacy deep-idle states. 
The proposed methodology offers the means for an early-stage design exploration and insights into idle time behavior and opportunities for varying server system configurations and load.
\end{abstract}

\section{Introduction}

Two of the main contributors to the power of a modern server are the processor's cores domain and uncore (package) ~\cite{barroso:dcc:2018}.
Existing power management techniques, such as sleep states~\cite{gough:energy-efficient-servers:book:2015} (i.e., C-states), target the cores and the package domains by exploiting their idle time.
Specifically, when a core (or all cores in a package) is idle, the system predicts the idle duration and transitions into an appropriate sleep state. Deeper sleep states provide greater power savings but incur higher performance overheads. The performance and power trade-offs 
of sleep states makes them unsuitable for certain types of applications~\cite{fowler:microservices:2014,sriraman:mu:iiswc:2018} with unpredictable active/idle periods~\cite{chou:DPM:HPCA:2019} and very strict (microsecond-scale) QoS constraints~\cite{sriraman:mu:iiswc:2018,zhan:CARB:CAL:2017} (i.e., microservices, real-time applications). As a result, datacenter operators often disable these features~\cite{keysight:cstates:2021,dell:cstates:2012,lenovo:cstates:2021}, undermining the server’s energy efficiency.

Prior research addresses some of the limitations of existing sleep-states
by fine-tuning sleep states to exploit latency slack 
~\cite{chou:DPM:HPCA:2019,chou:dynsleep:islped:2016}. Others focus on core consolidation 
~\cite{asyabi:peafowlsocc:2020,zhan:CARB:CAL:2017}. Additionally, some proposals investigate how to improve the accuracy of the idle time prediction (i.e., idle governor) through the use of machine learning~\cite{Sharafzadeh:yawn:apsys:2019,roba:idle-governor:iccp:2015}. Despite all these proposals showing improvement compared to the state of the art, there is limited analysis of the maximum theoretical idle opportunity, sources of inefficiency for exploiting idleness, and analysis of idleness for various applications and architectures.

We aim to fill this gap by developing an analytical model to estimate the idle time distribution for a core and a package under different service times and core count. To quantify the idle opportunity missed in existing servers, we conduct an experimental study using a synthetic workload that adheres to the constraints of our model, and estimate how much real idle measurements deviate from theoretical. Our analysis reveals that for both low (microseconds) and high service times (milliseconds), there is a lost idle opportunity whereby an idling system remains in a power-hungry active state instead of going to a deep idle state.


\section{Model Description}

The idle time distribution is modeled under the assumptions of the M/M/1, c$\times$M/M/1, and M/M/c queues. This section focuses primarily on M/M/1, as the core modeling principles apply similarly to c$\times$M/M/1 and M/M/c queues. 

\begin{figure}[t]
    \centering
    \includegraphics[width=1\linewidth]{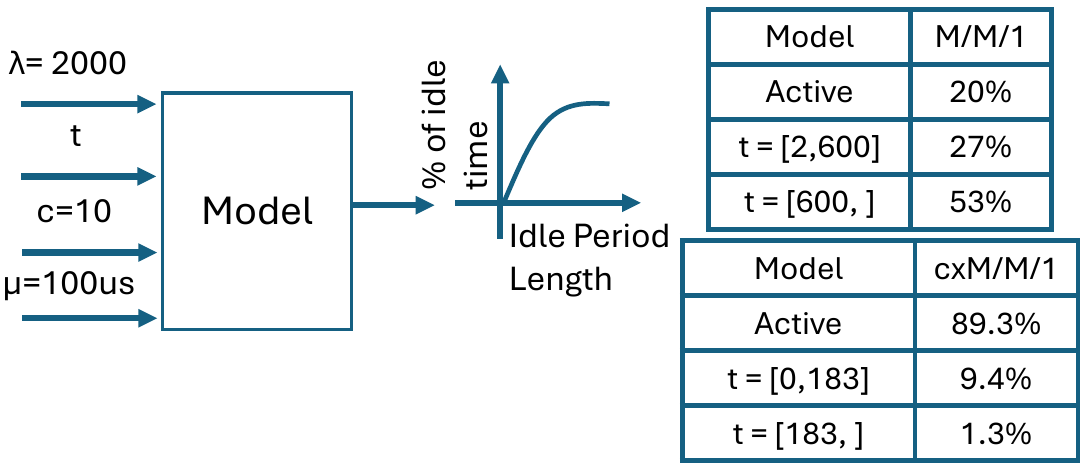}
    \caption{High-level description of model that estimates full idle time distribution for an M/M/1, c$x$M/M/1 and M/M/c queue model.}
    \label{fig:model-description}
\end{figure}

The \textbf{M/M/1} \cite{mor:queueing-theory:book:2013} 
arrivals follow a \emph{Poisson process}, service times follow an \emph{exponential distribution}, and there is a single server (i.e., \emph{c}=1). In this model, the idle period also follows an exponential distribution with rate $\lambda$, which corresponds to the inter-arrival time rate distribution of the system~\cite{mor:queueing-theory:book:2013,omahen:mmc:acm:1978}. Given inputs $\lambda$, \emph{t}, and \emph{c} (Fig.~\ref{fig:model-description}), the model integrates the idle distribution to produce the percentage of idle time lasting at least \emph{t} seconds for a system with \emph{c} cores. This output, along with the mean service time of the queuing model, are used to generate the percentage of time the system is active and the percentage of time the system is idle for a certain duration. 

Fig.~\ref{fig:model-description} shows an example of the model's inputs and outputs. For an arrival rate ($\lambda$) of 2000 (per core), and a core count (\emph{c}) of 10, the M/M/1 model estimates that a core will be idle for at least 2 to 600us for 27\% of the time. In the case of c$\times$M/M/1, the package will be idle for at most 183us for 9.4\% of the time. 

\section{Experimental Methodology}
\noindent\textbf{System.} 
Our baseline system is a 2-socket server with 2 Skylake-based (Intel Xeon Silver 4114) processors, sourced from CloudLab \cite{duplyakin:datacenterconstantischange:ccgrid:2020}.
 
\noindent\textbf{Workloads.} We use synthetic workloads to test the model on real systems. The programs have tunable service latency, number of threads, and seed, and emulate the behavior of an M/M/1, c$\times$M/M/1, and M/M/c queue. The processing time is implemented using a busy-wait loop. The client used is the modified version (DeathStar~\cite{deathstar} version) of the $wrk2$~\cite{wrk2} HTTP workload generator.  

\noindent\textbf{Measurement Tools.} We measure the core C-state residency using turbostat \cite{turbostat}, 
that reports frequency and idle statistics. We also use SoC Watch \cite{socwatch}
to measure the full package idle time and estimate package C-state residency. 
\section{Experimental Evaluation}\label{sec:evaluation}

\subsection{Core Idle Opportunity}\label{sec:evaluation:core-idleness}

\begin{figure}[t]
    \centering
    \includegraphics[width=1\linewidth]{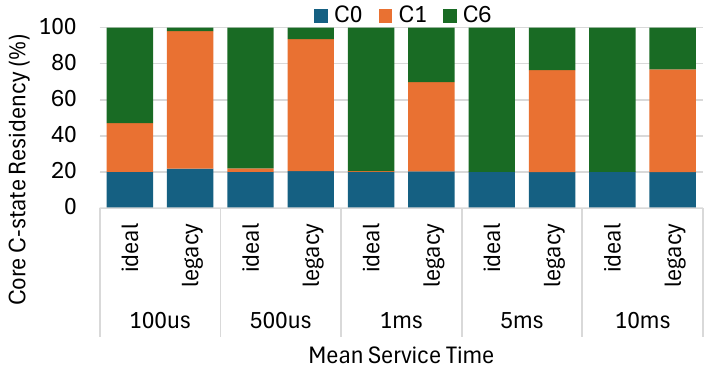}
    \caption{Core C-state (M/M/1) residency for an ideal and legacy system for mean service times: 100us, 500us, 1ms, 5ms, 10ms, at 20\% utilization.}
    \label{fig:core-idleness}
\end{figure}

We first measure the actual C-state residency (the percentage of time a core spends at a specific C-state) while running a synthetic workload that emulates an M/M/1 queue. We then estimate the theoretical idle-time distribution with our analytical model by setting the variable t to represent the overhead of a C-state transition. 
Fig. \ref{fig:core-idleness} compares the real measurements (legacy) with the model estimations (ideal)
for a utilization of 20\%. 

The analysis reveals that there is a significant lost idle opportunity on real legacy systems. 
%
The legacy system spends less time in deep sleep states compared to the ideal. Specifically, for 100us mean service time, the ideal system has a C6 residency of 53\%, whereas the legacy system achieves a core C6 residency of 2\%. These trends remain the same for higher mean service times, where for 10ms the maximum achievable C6 residency is 80\%, and the system achieves 23\%. This could be an indicator of a pessimistic hardware idle governor. 

\subsection{Package Idle Opportunity}\label{sec:evaluation:pkg-idleness}
We run a synthetic workload that emulates a c$\times$M/M/1 queue and capture the per core C-state trace of SoC Watch and analyze it to estimate the package C-state residency of the system for the following sleep states: 1) PC0 state which is the active state, 2) PC0-idle which is the state where all cores are idle but there are no power savings, and 3) PC6 which is a deep sleep state. We then compare the real measurements (legacy) with the results of our model (ideal) for the same mean service times. 

\begin{figure}[t]
    \centering
    \includegraphics[width=1\linewidth]{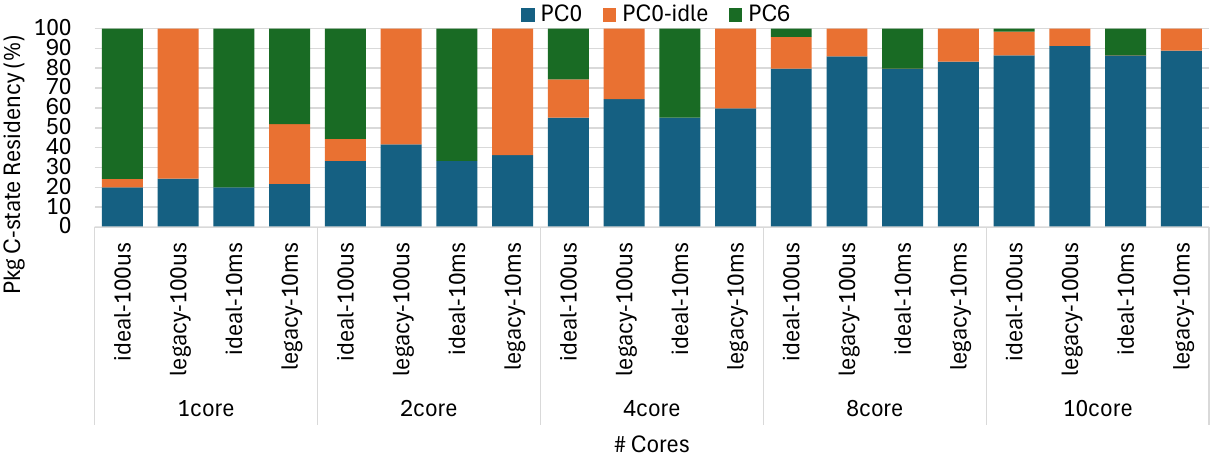}
    \caption{Pkg C-state (cxM/M/1) residency for an ideal and legacy system for 100us and 10ms mean service times at 20\% utilization.}
    \label{fig:pkg-residency}
\end{figure}

The actual system has an opportunity to enter package C-state for a much smaller fraction of the time (Fig.~\ref{fig:pkg-residency}). This is because PC6 has high transition overhead, and as a result, the system cannot always enter it. Another reason is that PC6 requires all cores of the package to be in a core C6 (i.e., CC6 C-state). Our previous analysis (see Sec.~\ref{sec:evaluation:core-idleness}) shows that a core in a legacy system spends most of it's time in shallow sleep states and thus reduces the opportunity for the package to enter PC6.


Another observation,
is that the full package idle opportunity gets smaller for higher core count for both low and high mean service time. This suggests that breaking packages into multiple domains of smaller core counts can possibly be more energy efficient. 
 

\section{Discussion}

\noindent\textbf{Core-Level Recommendations:} 
The reduced deep sleep state residency at the core level points to potential optimizations in the hardware/software stack that are responsible for transitioning in and out of sleep states. This could involve lightweight and accurate idle governors and/or optimized HW C-state flows. 

\noindent\textbf{Package-Level Recommendations:} At the package level, the analysis indicates that systems with fewer cores exhibit greater idle opportunity. This suggests that independent domains with lower core counts may offer a more energy-efficient design path.

\section{Conclusions}

In this work, we investigate the idle opportunity available at both the core and package levels in modern server processors using a model based on queuing theory. Our study reveals that there is unrealized idle power savings, as indicated by significantly increased potential for i) a core residing in C6 C-state, and ii) a package residing in PC6 (when all cores are idle). 
We believe that this model-based analysis provides useful insights into system idle behavior that will help shape appropriate power management-related features. At the same time, we believe that the developed model can be of valuable use as it can support early-stage design evaluation and can be extended to investigate different system characteristics, e.g., interrupt overheads. 
\section*{Acknowledgment}
This work is funded in part by the University of Cyprus research program RSHIELD.

\bibliographystyle{IEEEtran}
\bibliography{refs}

\begin{thebibliography}{10}
\providecommand{\url}[1]{#1}
\csname url@samestyle\endcsname
\providecommand{\newblock}{\relax}
\providecommand{\bibinfo}[2]{#2}
\providecommand{\BIBentrySTDinterwordspacing}{\spaceskip=0pt\relax}
\providecommand{\BIBentryALTinterwordstretchfactor}{4}
\providecommand{\BIBentryALTinterwordspacing}{\spaceskip=\fontdimen2\font plus
\BIBentryALTinterwordstretchfactor\fontdimen3\font minus \fontdimen4\font\relax}
\providecommand{\BIBforeignlanguage}[2]{{%
\expandafter\ifx\csname l@#1\endcsname\relax
\typeout{** WARNING: IEEEtran.bst: No hyphenation pattern has been}%
\typeout{** loaded for the language `#1'. Using the pattern for}%
\typeout{** the default language instead.}%
\else
\language=\csname l@#1\endcsname
\fi
#2}}
\providecommand{\BIBdecl}{\relax}
\BIBdecl

\bibitem{barroso:dcc:2018}
L.~A. Barroso, U.~Holzle, P.~Ranganathan, and M.~Martonosi, \emph{The Datacenter As a Computer: Designing Warehouse-scale Machines}, 3rd~ed.\hskip 1em plus 0.5em minus 0.4em\relax Morgan \& Claypool Publishers, 2018.

\bibitem{gough:energy-efficient-servers:book:2015}
C.~Gough, I.~Steiner, and W.~A. Saunders, \emph{Energy Efficient Servers: Blueprints for Data Center Optimization}, 1st~ed.\hskip 1em plus 0.5em minus 0.4em\relax USA: Apress, 2015.

\bibitem{fowler:microservices:2014}
{James Lewis and Martin Fowler}, ``Microservices: a definition of this new architectural term,'' online, \url{https://martinfowler.com/articles/microservices.html}, 2014.

\bibitem{sriraman:mu:iiswc:2018}
A.~Sriraman and T.~F. Wenisch, ``{$\mu$Suite: a Benchmark Suite for Microservices},'' in \emph{IISWC}, 2018.

\bibitem{chou:DPM:HPCA:2019}
C.-H. Chou, L.~N. Bhuyan, and D.~Wong, ``{$\mu$}dpm: Dynamic power management for the microsecond era,'' in \emph{2019 IEEE International Symposium on High Performance Computer Architecture (HPCA)}, 2019, pp. 120--132.

\bibitem{zhan:CARB:CAL:2017}
\BIBentryALTinterwordspacing
X.~Zhan, R.~Azimi, S.~Kanev, D.~Brooks, and S.~Reda, ``Carb: A c-state power management arbiter for latency-critical workloads,'' \emph{IEEE Comput. Archit. Lett.}, vol.~16, no.~1, p. 6–9, Jan. 2017. [Online]. Available: \url{https://doi.org/10.1109/LCA.2016.2537802}
\BIBentrySTDinterwordspacing

\bibitem{keysight:cstates:2021}
Keysight, ``{Performance Tuning Guide for Cisco UCS M5 Servers - White Paper},'' accessed Nov 2021, https://bit.ly/3nEq4CY.

\bibitem{dell:cstates:2012}
Dell, ``{BIOS Performance and Power Tuning Guidelines for Dell PowerEdge 12th Generation Servers},'' accessed Nov 2021, https://bit.ly/3llqoFh.

\bibitem{lenovo:cstates:2021}
Lenovo, ``{Tuning UEFI Settings for Performance and Energy Efficiency on Intel Xeon Scalable Processor-Based ThinkSystem Servers},'' accessed Nov 2021, https://lenovopress.com/lp1477.pdf.

\bibitem{chou:dynsleep:islped:2016}
\BIBentryALTinterwordspacing
C.-H. Chou, D.~Wong, and L.~N. Bhuyan, ``Dynsleep: Fine-grained power management for a latency-critical data center application,'' in \emph{Proceedings of the 2016 International Symposium on Low Power Electronics and Design}, ser. ISLPED '16.\hskip 1em plus 0.5em minus 0.4em\relax New York, NY, USA: Association for Computing Machinery, 2016, p. 212–217. [Online]. Available: \url{https://doi.org/10.1145/2934583.2934616}
\BIBentrySTDinterwordspacing

\bibitem{asyabi:peafowlsocc:2020}
\BIBentryALTinterwordspacing
E.~Asyabi, A.~Bestavros, E.~Sharafzadeh, and T.~Zhu, ``Peafowl: in-application cpu scheduling to reduce power consumption of in-memory key-value stores,'' in \emph{Proceedings of the 11th ACM Symposium on Cloud Computing}, ser. SoCC '20.\hskip 1em plus 0.5em minus 0.4em\relax New York, NY, USA: Association for Computing Machinery, 2020, p. 150–164. [Online]. Available: \url{https://doi.org/10.1145/3419111.3421298}
\BIBentrySTDinterwordspacing

\bibitem{Sharafzadeh:yawn:apsys:2019}
\BIBentryALTinterwordspacing
E.~Sharafzadeh, S.~A.~S. Kohroudi, E.~Asyabi, and M.~Sharifi, ``Yawn: A cpu idle-state governor for datacenter applications,'' in \emph{Proceedings of the 10th ACM SIGOPS Asia-Pacific Workshop on Systems}, ser. APSys '19.\hskip 1em plus 0.5em minus 0.4em\relax New York, NY, USA: Association for Computing Machinery, 2019, p. 91–98. [Online]. Available: \url{https://doi.org/10.1145/3343737.3343740}
\BIBentrySTDinterwordspacing

\bibitem{roba:idle-governor:iccp:2015}
A.~Roba and Z.~Baruch, ``An enhanced approach to dynamic power management for the linux cpuidle subsystem,'' in \emph{2015 IEEE International Conference on Intelligent Computer Communication and Processing (ICCP)}, 2015, pp. 511--517.

\bibitem{mor:queueing-theory:book:2013}
M.~Harchol-Balter, \emph{Performance Modeling and Design of Computer Systems: Queueing Theory in Action}, 1st~ed.\hskip 1em plus 0.5em minus 0.4em\relax USA: Cambridge University Press, 2013.

\bibitem{omahen:mmc:acm:1978}
\BIBentryALTinterwordspacing
K.~Omahen and V.~Marathe, ``Analysis and applications of the delay cycle for the m/m/c queueing system,'' \emph{J. ACM}, vol.~25, no.~2, p. 283–303, Apr. 1978. [Online]. Available: \url{https://doi.org/10.1145/322063.322072}
\BIBentrySTDinterwordspacing

\bibitem{duplyakin:datacenterconstantischange:ccgrid:2020}
\BIBentryALTinterwordspacing
D.~Duplyakin, A.~Uta, A.~Maricq, and R.~Ricci, ``In datacenter performance, the only constant is change,'' in \emph{2020 20th IEEE/ACM International Symposium on Cluster, Cloud and Internet Computing (CCGRID)}.\hskip 1em plus 0.5em minus 0.4em\relax Los Alamitos, CA, USA: IEEE Computer Society, may 2020, pp. 370--379. [Online]. Available: \url{https://doi.ieeecomputersociety.org/10.1109/CCGrid49817.2020.00-56}
\BIBentrySTDinterwordspacing

\bibitem{deathstar}
\BIBentryALTinterwordspacing
Y.~Gan, Y.~Zhang, D.~Cheng, A.~Shetty, P.~Rathi, N.~Katarki, A.~Bruno, J.~Hu, B.~Ritchken, B.~Jackson, K.~Hu, M.~Pancholi, Y.~He, B.~Clancy, C.~Colen, F.~Wen, C.~Leung, S.~Wang, L.~Zaruvinsky, M.~Espinosa, R.~Lin, Z.~Liu, J.~Padilla, and C.~Delimitrou, ``An open-source benchmark suite for microservices and their hardware-software implications for cloud \& edge systems,'' in \emph{Proceedings of the Twenty-Fourth International Conference on Architectural Support for Programming Languages and Operating Systems}, ser. ASPLOS '19.\hskip 1em plus 0.5em minus 0.4em\relax New York, NY, USA: Association for Computing Machinery, 2019, p. 3–18. [Online]. Available: \url{https://doi.org/10.1145/3297858.3304013}
\BIBentrySTDinterwordspacing

\bibitem{wrk2}
G.~Tene, ``wrk2 - a constant throughput, correct latency recording variant of wrk,'' \url{https://github.com/giltene/wrk2}, 2015, accessed: 2025-05-09.

\bibitem{turbostat}
L.~Brown, ``turbostat: Report processor frequency and idle stats,'' \url{https://www.kernel.org/doc/html/latest/tools/power/turbostat.html}, 2024, accessed: 2025-05-09.

\bibitem{socwatch}
{Intel}, ``{{Energy Analysis User Guide - SoC Watch}},'' https://intel.ly/3jZemQI.

\end{thebibliography}

\end{document}